\documentclass[12pt,a4paper]{article}
\usepackage{amssymb,amsmath}
\usepackage[dvips]{lscape,graphicx}

\newcommand{\bc}{\begin{center}}
\newcommand{\ec}{\end{center}}
\newcommand{\bd}{\begin{displaymath}}
\newcommand{\ed}{\end{displaymath}}
\newcommand{\be}{\begin{equation}}
\newcommand{\ee}{\end{equation}}
\newcommand{\ba}{\begin{array}}
\newcommand{\ea}{\end{array}}
\newcommand{\bt}{\begin{tabular}}
\newcommand{\et}{\end{tabular}}

\newcommand{\ds}{\displaystyle}

\begin{document}

\title{LHC signatures of neutral pseudo--Goldstone boson\\ in the E$_6$CHM}

\author{R.~Nevzorov\footnote{On leave of absence from the Theory Department, SSC RF ITEP of NRC "Kurchatov Institute", Moscow, Russia.},
A.~W.~Thomas\qquad\qquad\\[5mm]
\itshape{ARC Centre of Excellence for Particle Physics at the Terascale and CSSM,}\\[0mm]
\itshape{Department of Physics, The University of Adelaide, Adelaide SA 5005, Australia}}

\date{}

\maketitle

\begin{abstract}{
\noindent
The breakdown of the $SU(6)$ global symmetry to its $SU(5)$ subgroup, that contains the
standard model (SM) gauge group, in the $E_6$ inspired composite Higgs model (E$_6$CHM)
results in a set of pseudo--Nambu--Goldstone bosons (pNGBs). This set, in particular,
involves the SM--like Higgs doublet and a SM singlet boson.  In the limit when CP is conserved
the SM singlet scalar $A$ is a CP--odd state that does not mix with the SM--like Higgs.
The interactions of $A$ with exotic matter beyond the SM, which ensures anomaly
cancellation and approximate gauge coupling unification, can induce couplings of this
pseudoscalar to the SM gauge bosons. We specify the interactions of the SM singlet pNGB state
with the exotic vector--like fermions, top quark and SM gauge bosons. Also we explore the
dependence of the branching ratios of the pseudoscalar $A$ and its LHC production cross section
on the parameters of the E$_6$CHM.
}
\end{abstract}

\newpage
\section{Introduction}
After the discovery of a scalar particle, which is consistent with the Standard Model (SM) Higgs boson,
at the LHC Run-1, the focus of experimental and theoretical studies is shifting again towards the investigations
of possible new physics phenomena beyond the SM (BSM). An amount of data at the $13\,\mbox{TeV}$ LHC
collected by ATLAS and CMS already permits us to set competitive constraints on the parameters of some
BSM models as compared with those obtained at $8\,\mbox{TeV}$. In this article we consider the possible
collider signatures associated with the presence of the neutral pseudo--Goldstone boson in the framework of the
$E_6$ inspired composite Higgs model (E$_6$CHM) \cite{Nevzorov:2015sha}. As in any other composite
Higgs model \cite{Agashe:2004rs}  (for a recent review, see \cite{Bellazzini:2014yua}), the E$_6$CHM involves weakly--coupled elementary
and strongly interacting sectors. The E$_6$SSM implies that at some high energy scale, $M_X$, the $E_6\times G_0$ gauge group is broken
down to the $SU(3)_C\times SU(2)_W\times U(1)_Y \times G$ subgroup where $G_0$ and $G$ are associated with the strongly interacting
sector and $SU(3)_C\times SU(2)_W\times U(1)_Y$ is the SM gauge group. Fields belonging to the strongly coupled sector can be
charged under both the $E_6$ and $G_0$ ($G$) gauge symmetries. The weakly--coupled sector contains elementary states that
participate in the $E_6$ interactions only. In the E$_6$CHM the appropriate suppression of the proton decay rate and the Majorana
masses of the left--handed neutrino can be achieved if global $U(1)_B$ and $U(1)_L$ symmetries, which ensure the conservation
of baryon and lepton numbers, are imposed. Because of the almost exact conservation of the $U(1)_B$ and $U(1)_L$ charges,
the elementary states with different baryon and/or lepton numbers should stem from different $27$--plets, while all other components
of these $27$--plets have to gain masses of the order of $M_X$.

The appropriate splitting of the $E_6$ fundamental representations may occur within a six--dimensional orbifold GUT model with
$N=1$ supersymmetry (SUSY) \cite{Nevzorov:2015sha} in which SUSY is broken slightly below the GUT scale $M_X$\footnote{Different
phenomenological aspects of the $E_6$ inspired models with low-scale supersymmetry breaking were recently studied in \cite{King:2016wep}-\cite{e6ssm}.}.
In this model the SM fermions with different baryon and/or lepton numbers are components of different bulk $27$--plets. All fields from the
strongly interacting sector are localised on the brane, where $E_6$ symmetry is broken to the $SU(6)\times SU(2)_N$ subgroup.
It is assumed that the $E_6$ gauge symmetry gets broken to the SM gauge group, $SU(2)_N$ symmetry is entirely broken,
while $SU(6)$, which contains the $SU(3)_C\times SU(2)_W\times U(1)_Y$ subgroup, remains an approximate global symmetry of the
composite sector at low energies.

Below the scale $f$ ($f\gtrsim 10\,\mbox{TeV}$) the global $SU(6)$ symmetry in the E$_6$CHM is
expected to be broken down to $SU(5)$, so that the SM gauge group is preserved, leading to eleven pseudo--Nambu--Goldstone
bosons (pNGBs). One of these pNGB states is a SM singlet field, four others form the SM--like Higgs doublet, $H$, and six
pNGB states are associated with an $SU(3)_C$ triplet, $T$. None of these pNGB states carry any baryon and/or lepton numbers.
The effective potential, which describes the interactions of the pNGB states, is induced by radiative corrections. These corrections
are caused by the interactions between the elementary states and their composite partners that violate $SU(6)$ symmetry.
The induced effective potential has a structure that results in the spontaneous breakdown of the electroweak (EW) symmetry,
while $SU(3)_C$ remains intact. In this case the effective quartic Higgs coupling tends to be small and, therefore, may lead to
a $125\,\mbox{GeV}$ Higgs.

Although the SM gauge couplings $\alpha_i (M_X)$ in the orbifold GUT models may not be identical an approximate unification
of these couplings is expected to take place near the scale $M_X$. Such a unification can be attained if the right--handed top
quark $t^c$ is entirely composite and the weakly--coupled sector, together with the SM fields (but without $t^c$), involves a set
of exotic states. The presence of additional exotic states also guarantees the cancellation of gauge anomalies in the elementary
sector. In the E$_6$CHM the exotic states mentioned above get combined with the composite counterparts, which fill complete
$SU(5)$ representations, resulting in a set of vector--like fermion states and composite $t^c$. In this article we argue that the
interactions of these vector--like fermions with the SM singlet pNGB state can lead to the interesting collider signatures that
can be observed at the LHC in the near future.

The layout of this article is as follows. In the next Section we discuss the matter content of the weakly--coupled elementary
sector, the generation of masses of the SM particles, unification of the SM gauge couplings and non--linear realization of
the Higgs mechanism in the E$_6$CHM. The interactions of the exotic elementary states with their composite partners is
also considered. In Section 3 we specify the interactions of the SM singlet pNGB state with the exotic vector--like fermions,
top quark and SM gauge bosons. In section 4 we explore the possibility of obtaining experimental evidence to support the
model proposed here by examining the branching ratios and the LHC production cross section associated with this pNGB state.
Section 5 concludes the paper.

\section{The E$_6$CHM}

\subsection{Gauge coupling unification and $SU(6)$ symmetry breaking}

As already mentioned, approximate gauge coupling unification can be achieved in the E$_6$CHM.
This scenario is realised if the right--handed top quark, $t^c$, is composite and the sector of weakly--coupled elementary
states contains the following set of matter multiplets (see also \cite{Agashe:2005vg}):
\begin{equation}
(q_i,\,d^c_i,\,\ell_i,\,e^c_i) + u^c_{\alpha} + \bar{q}+\bar{d^c}+\bar{\ell}+\bar{e^c}+\eta\,,
\label{1}
\end{equation}
where $\alpha=1,2$ runs over the first two generations and $i=1,2,3$ runs over all three.
In Eq.~(\ref{1}) $q_i$ and $\ell_i$ represent left-handed quark and lepton doublets,
$u_{\alpha}^c, d_i^c$ and $e_i^c$ correspond to the right-handed up- and down-type quarks and charged leptons,
while $\bar{q},\,\bar{d^c},\,\bar{\ell}$ and $\bar{e^c}$ are associated with the exotic states that have exactly opposite
$SU(3)_C\times SU(2)_W\times U(1)_Y$ quantum numbers to the left-handed quark doublets, right-handed down-type quarks,
left-handed lepton doublets and right-handed charged leptons, respectively. The set of fermion states (\ref{1}) is chosen so
that the weakly--coupled elementary sector involves all SM fermions except $t^c$ and anomaly cancellation takes place.

An extra exotic state, $\eta$, with spin $1/2$, which is also included in the set
of states (1), does not participate in the $SU(3)_C\times SU(2)_W\times U(1)_Y$
gauge interactions. It is introduced to ensure the phenomenological viability of
this model.  As explained in the next subsection and Section 3, the
elementary exotic fermion states mentioned above get combined with their
composite partners forming a set of vector--like Dirac fermions. The lightest of
these Dirac states tends to be stable. The phenomenological viability of the model
under consideration implies that the lightest exotic fermion state is neutral, does
not participate in the strong interactions and has rather suppressed coupling to
the $Z$-boson. Thus it should be predominantly a superposition of $\eta$ and
its composite partner. Then this state can also serve as a dark matter candidate
if its mass is of the order of a few hundred GeV.

At low energies ($E\lesssim 4\pi f$) the strongly interacting sector in the composite Higgs models \cite{Agashe:2004rs} leads
to a set of bound states that includes the pNGB states as well as massive fields with quantum numbers of all SM particles.
These are the so--called composite partners of  the SM fermions and bosons. The contributions of these new states
to electroweak precision observables were analysed in Refs. \cite{EWPOCHM}--\cite{Vignaroli:2012si}.
In the  E$_6$CHM the composite bound states fill complete $SU(6)$ representations, that involve the composite partners
of quarks, leptons and gauge bosons. The elementary states couple to the appropriate operators of the strongly interacting
sector that give rise to the mixing between these states and their composite partners. The partial compositeness of
the SM bosons and fermions makes possible the generation of their masses
caused by non-zero vacuum expectation value (VEV) of the pNGB state associated with the Higgs boson\footnote{It is expected
that nonperturbative effects in the ultraviolate (UV) completions
of the composite Higgs models should induce the breakdown of global symmetry,
that results in a set of the pNGB states including the Higgs doublet, as well as
leading to the mixing between elementary states and their composite partners, which
gives rise to all masses and mixing in the quark and lepton sectors. The construction
of such UV completions is a challenging problem, which is beyond the scope of
the present paper. Here we just assume that the gauge groups $G_0$ and $G$
can be chosen so that a phenomenologically viable model of this type can be
constructed.}. The couplings
of the SM states to the composite Higgs are determined by the fractions of the compositeness of these states.
In most cases, especially for the first and second generations of fermions, the corresponding fractions are sufficiently small.
Therefore the non-diagonal flavor transitions and the modifications of the $W$ and $Z$ couplings associated with the
light SM fermions are suppressed. Within the composite Higgs models the constraints that come from flavour-changing
processes in the quark and lepton sectors were examined in Refs. \cite{Barbieri:2012tu}--\cite{Barbieri:2012uh} and
\cite{Barbieri:2012uh}--\cite{Csaki:2008qq}, respectively. In particular, it was shown that in the case when the matrices of
effective Yukawa couplings in the strong sector are structureless, i.e anarchic matrices, adequate
suppression of flavor changing neutral currents (FCNCs) can be obtained only if $f$ is larger than $10\,\mbox{TeV}$
\cite{Barbieri:2012tu}--\cite{Csaki:2008zd}, \cite{Redi:2011zi}--\cite{Blanke:2008zb}, \cite{Agashe:2006iy}\footnote{This
bound on the scale $f$ can be significantly alleviated in the composite Higgs models with
flavour symmetries \cite{Barbieri:2008zt}--\cite{Barbieri:2012tu}, \cite{Redi:2011zi}, \cite{Barbieri:2012uh}--\cite{Redi:2013pga},
\cite{Cacciapaglia:2007fw}.}.   The implications of the composite Higgs models were also
studied for Higgs physics \cite{Bellazzini:2012tv}--\cite{Azatov:2013ura}, \cite{Mrazek:2011iu}--\cite{Pomarol:2012qf},
gauge coupling unification \cite{Gherghetta:2004sq}--\cite{Barnard:2014tla}, dark matter \cite{Frigerio:2011zg}, \cite{Frigerio:2012uc},
\cite{Barnard:2014tla}--\cite{Asano:2014wra} and collider phenomenology \cite{Pomarol:2008bh}--\cite{Bellazzini:2012tv},
\cite{Barbieri:2008zt}, \cite{Redi:2011zi}, \cite{Redi:2013pga}, \cite{Pomarol:2012qf}, \cite{Delaunay:2013pwa}.
Non--minimal composite Higgs models were considered in Refs. \cite{Frigerio:2011zg}, \cite{Mrazek:2011iu}--\cite{Frigerio:2012uc},
\cite{Barnard:2014tla}--\cite{Asano:2014wra}, \cite{Cacciapaglia:2014uja}.

The presence of additional exotic states in the E$_6$CHM facilitates the convergence of the SM gauge couplings at very high energies.
Indeed, all states in the strongly coupled sector come in complete $SU(6)$ and $SU(5)$ representations which contribute equally to
the one--loop beta functions of the $SU(3)_C$, $SU(2)_W$ and $U(1)_Y$ interactions. Thus composite sector fields should not spoil
the convergence of the SM gauge couplings in the leading approximation, which is determined by the matter content of the elementary
sector. Using the one--loop renormalisation group equations (RGEs) one can find that for $\alpha(M_Z)=1/127.9$, $\sin^2\theta_W=0.231$
and the elementary particle spectrum given by Eq.~(\ref{1}), the exact unification of the SM gauge couplings takes place if
$\alpha_3(M_Z)\simeq 0.109$\,. The corresponding gauge coupling unification scale is somewhat close to
$M_X\sim 10^{15}-10^{16}\, \mbox{GeV}$. Although $\alpha_3(M_Z)\simeq 0.109$ is considerably smaller than the
central measured low energy value of this coupling, this estimation demonstrates that an approximate gauge coupling unification may
be achieved in the E$_6$CHM. Moreover, it was also argued that the inclusion of higher order effects may improve the unification of the
SM gauge couplings \cite{Agashe:2005vg}, \cite{Barnard:2014tla}.

In the E$_6$CHM the global $SU(6)$ symmetry of the strongly interacting sector is broken down to $SU(5)$ below the scale $f$.
We denote the generators of the $SU(5)$ subgroup of $SU(6)$ by $T^a$, whereas the eleven generators from the coset $SU(6)/SU(5)$
associated with the pNGB states are denoted by $T^{\hat{a}}$. Here the $SU(6)$ generators are normalised so that
$\mbox{Tr} T^a T^b = \ds\frac{1}{2} \delta_{ab}$. It is convenient to use the non--linear representation of the pNGB states
in terms of a 6--component unit vector $\Omega$, which is a fundamental representation of $SU(6)$, that is \cite{Nevzorov:2015sha}
\be
\ba{c}
\Omega^T = \Omega_0^T \Sigma^T = e^{i\frac{\phi_0}{\sqrt{15}f}}
\Biggl(C \phi_1\quad C \phi_2\quad C \phi_3\quad C \phi_4\quad C\phi_5\quad \cos\dfrac{\tilde{\phi}}{\sqrt{2} f} + \sqrt{\dfrac{3}{10}} C \phi_0 \Biggr)\,,\\[3mm]
C=\dfrac{i}{\tilde{\phi}} \sin \dfrac{\tilde{\phi}}{\sqrt{2} f}\,,\qquad \tilde{\phi}=\sqrt{\dfrac{3}{10}\phi_0^2+|\phi_1|^2+|\phi_2|^2+|\phi_3|^2+|\phi_4|^2+|\phi_5|^2}\,,
\ea
\label{2}
\ee
where
$$
\Omega_0^T= (0\quad 0\quad 0\quad 0\quad 0\quad 1)\,,\qquad \Sigma= e^{i\Pi/f}\,,\qquad \Pi=\Pi^{\hat{a}} T^{\hat{a}}\,.
$$
The fields $\phi_1\,, \phi_2\,, \phi_3\,, \phi_4$ and $\phi_5$ are complex while $\phi_0$ is a real field. Vector $\Omega$ transforms
as $\bf{5}+\bf{1}$ under the transformation of the unbroken $SU(5)$ subgroup where $\bf{5}=\tilde{H}\sim (\phi_1\,\, \phi_2\,\, \phi_3\,\, \phi_4\,\, \phi_5)$
and $\bf{1}=\phi_0$ is a SM singlet field.
The first two components of $\tilde{H}$ transform as an $SU(2)_W$ doublet, $H\sim (\phi_1\,\, \phi_2)$, and correspond to the SM--like Higgs doublet.
Three other components of $\tilde{H}$, $T\sim (\phi_3\,\, \phi_4\,\, \phi_5)$, are associated with the $SU(3)_C$ triplet. In the E$_6$CHM the components
of vector $\Omega$ do not carry any baryon and/or lepton numbers. In the leading approximation the Lagrangian, that describes their interactions,
is given by
\be
\mathcal{L}_{pNGB}=\ds\dfrac{f^2}{2}\biggl|\mathcal{D}_{\mu} \Omega \biggr|^2\,.
\label{3}
\ee

The pNGB effective potential $V_{eff}(\tilde{H}, T, \phi_0)$ can be obtained by integrating out the exotic fermions and heavy resonances of
the composite sector. It is induced by the interactions of the elementary fermions and gauge bosons with their composite partners,
that break $SU(6)$ global symmetry and vanish in the exact $SU(6)$ symmetry limit. The analysis of the pNGB effective potentials in
the composite Higgs models, which are similar to the E$_6$CHM, revealed that there is a considerable part of the parameter space
where the $SU(2)_W\times U(1)_Y$ gauge symmetry is broken to $U(1)_{em}$, associated with electromagnetism, while $SU(3)_C$
remains intact \cite{Frigerio:2011zg}, \cite{Barnard:2014tla}. Since in the E$_6$CHM the scale $f\gtrsim 10\,\mbox{TeV}$, a significant
tuning, $\sim 0.01\%$, is needed to get the quadratic term $m^2_H |H|^2$ in $V_{eff}(\tilde{H}, T, \phi_0)$ with the appropriate value
of the parameter $m^2_H $ that leads to a $125\,\mbox{GeV}$ Higgs state. It was shown that such tuning can be accomplished by cancelling
two different contributions to $m^2_H $ associated with the gauge fields and exotic fermions which appear with different signs
in such composite Higgs models \cite{Barnard:2014tla}. In these models the $SU(3)_C$ triplet scalar $T$ tends to be substantially
heavier than the SM--like Higgs boson.

\subsection{Exotic fermion states}

The phenomenological viability of the scenario under consideration implies that the dynamics
of the strongly coupled sector below the $SU(6)$ breaking scale $f$ results in the composite ${\bf 10} + {\bf \overline{5}} + {\bf 1}$ multiplets
of $SU(5)$. The components of these $SU(5)$ multiplets decompose under $SU(3)_C\times SU(2)_W\times U(1)_Y\times U(1)_B \times U(1)_L$
as follows:
\begin{equation}
\ba{rcl}
{\bf 10} &\to& Q=(U,\,D) = \left(3,\,2,\,\ds\frac{1}{6},\,-\ds\frac{1}{3},\,0\right)\,,\\[2mm]
&& t^c = \left(3^{*},\,1,\,-\ds\frac{2}{3},\,-\ds\frac{1}{3},\,0\right)\,,\\[2mm]
&& E^c = \left(1,\,1,\,1,\,-\ds\frac{1}{3},\,0\right)\,;\\[4mm]
{\bf \overline{5}} &\to & D^c = \left(\bar{3},\,1,\,\ds\frac{1}{3},\, \pm \ds\frac{1}{3},\,0 \right)\,,\\[2mm]
& & L=(N,\, E) = \left(1,\,2,\,-\ds\frac{1}{2},\, \pm \ds\frac{1}{3},\,0 \right)\,;\\[4mm]
{\bf 1} &\to & \bar{\eta}=\left(1,\,1,\,0,\, \mp\ds\frac{1}{3},\,0 \right)\,.
\ea
\label{4}
\end{equation}
The first and second quantities in brackets are the $SU(3)_C$ and $SU(2)_W$ representations, while the third, fourth and fifth
quantities are $U(1)_Y$, $U(1)_{B}$ and $U(1)_{L}$ charges respectively. The conservation of the $U(1)_B$ and $U(1)_L$ charges
implies that all components of the $10$--plet specified above carry the same baryon and lepton numbers as $t^c$.
The multiplets ${\bf \overline{5}}$ and ${\bf 1}$ are allowed to have baryon charges $-1/3$ and $+1/3$ \cite{Nevzorov:2015sha}.

The composite $SU(5)$ multiplets (\ref{4}) are expected to get combined with the elementary exotic states $\bar{q}$,\,
$\bar{e^c}$,\,$\bar{d^c}$,\,$\bar{\ell}$ and $\eta$ resulting in a set of vector--like fermion states. The only exceptions are
the components of the $10$--plet associated with the composite $t^c$, which survive down to the EW scale. In the E$_6$CHM the
exotic states $\bar{q}$,\, $\bar{e^c}$,\,$\bar{d^c}$,\,$\bar{\ell}$ and $\eta$ constitute the following incomplete multiplets
of $SU(6)$ at low energies
\be
\bar{q} \in {\bf \overline{15}}^q\,,\qquad \bar{e^c}\in {\bf \overline{15}}^e\,,\qquad \bar{d^c} \in {\bf 6}^d\,, \qquad
\bar{\ell}\in {\bf 6}^{\ell}\,,\qquad \eta\in {\bf 1}^{\eta}\,.
\label{5}
\ee
As discussed in Ref. \cite{Nevzorov:2015sha}, the composite partners of the up type quarks can stem from either
the totally antisymmetric third--rank tensor ${\bf 20}$ or the antisymmetric second--rank tensor ${\bf 15}$. These $SU(6)$
representations decompose under unbroken $SU(5)$ as follows: ${\bf 15}={\bf 10} \oplus {\bf 5}$ and
${\bf 20}={\bf 10} \oplus {\bf \overline{10}}$. Thus the $10$--plet associated with the composite $t^c$ may belong to
either a ${\bf 15}$--plet or ${\bf 20}$--plet of $SU(6)$. Further, we assume that this $10$--plet is a linear superposition of
the corresponding components of the ${\bf 15}$--plet (${\bf 15}^t$) and ${\bf 20}$--plet (${\bf 20}^t$). On the other hand,
the composite ${\bf \overline{5}}$ can originate from either the ${\bf \overline{15}}$--plet (${\bf \overline{15}}'$) or
${\bf \overline{6}}$--plet (${\bf \overline{6}}'$). Therefore it seems to be natural to expect that the composite
${\bf \overline{5}}$, which survives below scale $f$, is a superposition of the appropriate components of  ${\bf \overline{15}}'$
and ${\bf \overline{6}}'$. Here we also assume that the composite SM singlet state, $\bar{\eta}$, that arises below the scale $f$, is a linear
combination of ${\bf 1}$ and the corresponding component of ${\bf \overline{6}}$ of $SU(6)$. In the most general case,
the interactions of the $SU(6)$ composite multiplets mentioned above with the incomplete $SU(6)$ representations that involve
elementary exotic states (\ref{5}) can be written as
\be
\ba{rcl}
\mathcal{L}_{exotic} & = & \tilde{\mu}_{Q} {\bf \overline{15}}^q {\bf 15}^t +
\tilde{\sigma}_Q f {\bf \overline{15}}^q \Omega^{\dagger} {\bf 20}^t +
\tilde{\mu}_{E} {\bf \overline{15}}^e  {\bf 15}^t +
\tilde{\sigma}_{E} f {\bf \overline{15}}^e \Omega^{\dagger} {\bf 20}^t +
\tilde{\mu}_D {\bf \overline{6}}' {\bf 6}^d \\[2mm]
& + & \tilde{\sigma}_{D} f {\bf \overline{15}}' \Omega {\bf 6}^d +
\tilde{\mu}_L {\bf \overline{6}}' {\bf 6}^{\ell} +
\tilde{\sigma}_L f {\bf \overline{15}}' \Omega {\bf 6}^{\ell} +
\tilde{\mu}_{\eta} {\bf 1} {\bf 1}^{\eta} +
\tilde{\sigma}_{\eta} f {\bf \overline{6}} \Omega {\bf 1}^{\eta} + h.c.\,.
\ea
\label{6}
\ee

\section{Couplings of the SM singlet pNGB state}

Using Eqs.~(\ref{2}) and (\ref{6}) one can obtain the explicit analytical expressions for the masses of exotic fermion states
and their couplings to the SM singlet field $\phi_0=A$.  Here, motivated by the nonobservation of CP violation beyond the SM,
invariance under CP transformation is imposed. This forbids the mixing between $A$ and the SM--like Higgs state. Indeed, if all
couplings in Eq.~(\ref{6}) are real, then $A$ manifests itself in the Yukawa interactions with fermions as a pseudoscalar field.
As a consequence $A$ can not mix with the Higgs boson because of the almost exact CP--conservation.

In the leading approximation, the Lagrangian that describes the interactions between $A$
and other states is given by
$$
\mathcal{L}_{A}= \ds\frac{y_t}{\Lambda_t} A (i\bar{t}_L H t_R + h.c.) +
A \Biggl(i\kappa_D  \bar{d^c} D^c + i \kappa_Q \bar{q} Q + i \lambda_L  \bar{\ell} L + i \lambda_E  \bar{e^c} E^c +
i\lambda_{\eta} \bar{\eta}\eta +h. c. \Biggr)
$$
\begin{equation}
+ \ds\frac{\alpha_Y}{16\pi \Lambda_1} A B_{\mu\nu} \widetilde{B}^{\mu\nu}
+ \ds\frac{\alpha_2}{16\pi \Lambda_2} A W^a_{\mu\nu} \widetilde{W}^{a\mu\nu}
+ \ds\frac{\alpha_3}{16\pi \Lambda_3} A G^{\sigma}_{\mu\nu} \widetilde{G}^{\sigma\mu\nu}\,,
\label{7}
\end{equation}
where $B_{\mu\nu}$, $W^a_{\mu\nu}$, $G^{\sigma}_{\mu\nu}$ are field strengths for the $U(1)_Y$, $SU(2)_W$ and $SU(3)_C$
gauge interactions, respectively, whereas $\widetilde{G}^{\sigma\mu\nu}=\frac{1}{2}\epsilon^{\mu\nu\lambda\rho} G^{\sigma}_{\lambda\rho}$,
$\widetilde{W}^{a\mu\nu}=\frac{1}{2}\epsilon^{\mu\nu\lambda\rho} W^{a}_{\lambda\rho}$ and
$\widetilde{B}^{\mu\nu}=\frac{1}{2}\epsilon^{\mu\nu\lambda\rho} B_{\lambda\rho}$. Here $\alpha_Y=3\alpha_1/5$ while
$\alpha_1$, $\alpha_2$ and $\alpha_3$ are (GUT normalised) gauge couplings of $U(1)_Y$, $SU(2)_W$ and $SU(3)_C$ interactions.

In Eq.~(\ref{7}) all interactions between the pseudoscalar $A$ and SM fermions except the coupling of $A$ to the top quarks
were ignored, because the masses of the SM quarks and leptons are negligibly small as compared with the scale $f$ and the top quark
mass $m_t$. The first term in Eq.~(7) stems from the interactions
\be
\mathcal{L}_{t}= g_t {\bf 15}(Q) \Omega {\bf 20}^t + \tilde{g}_t {\bf 20}(Q) \Omega {\bf 15}^t + h. c.
\label{8}
\ee
in the strongly interacting sector, where the linear combination of the appropriate components of the $10$--plets of $SU(5)$ from
${\bf 15}(Q)$ and ${\bf 20}(Q)$ corresponds to the composite partners of the third generation left--handed quark doublet.
Then the mixing between the third generation left--handed quark doublet from the weakly--coupled elementary sector and
its composite partners leads to the first term in Eq.~(\ref{7}). In the case when $t^c$ is predominantly the appropriate
component of ${\bf 20}^t$ of $SU(6)$, the scale $\Lambda_t=\sqrt{15} f$. If $t^c$ is mainly a component of ${\bf 15}^t$
of $SU(6)$ then $\Lambda_t=\sqrt{\frac{60}{49}} f\simeq 1.1 f$.

The second term in Eq.~(\ref{7}) originates from the Lagrangian (\ref{6}). The interactions specified in Eq.~(\ref{6})
also give rise to the mass terms of the exotic fermions
\be
\mathcal{L}_{mass}= \mu_D \bar{d^c} D^c + \mu_Q \bar{q} Q + \mu_L  \bar{\ell} L +
\mu_E  \bar{e^c} E^c + \mu_{\eta} \bar{\eta}\eta +h.c.\,.
\label{9}
\ee
The masses $\mu_i$ in Eq.~(\ref{9}) are linear combinations of two contributions. One of these contributions is
proportional to $\tilde{\mu}_i$\footnote{In general the parameters $\tilde{\mu}_i$ as well as other mass
parameters that correspond to the mixing between elementary states and their composite
partners get induced by nonperturbative effects that may also give rise to
the breakdown of the $SU(6)$ global symmetry. Therefore one can expect that
in the simplest models $\tilde{\mu}_i$ should be of order of $f$. The precise
values of these mass parameters as well as the values of dimensionless
couplings $\tilde{\sigma}_i$ should depend on the ultraviolate completion of
the E$_6$CHM which is not specified in this article.} whereas another is induced as a result of the
breakdown of $SU(6)$ global symmetry to its $SU(5)$ subgroup, that contains the $SU(3)_C\times SU(2)_W\times U(1)_Y$
gauge symmetry and is therefore proportional to $\tilde{\sigma}_i f$. The approximate gauge coupling unification
requires exotic fermions to be substantially lighter than $10\,\mbox{TeV}$. This can be always achieved
by adjusting the mass parameters $\tilde{\mu}_i$ in Eq.~(\ref{6}). On the other hand, the Yukawa couplings
$\kappa_i$ and $\lambda_i$ in Eq.~(\ref{7}) are proportional to $\tilde{\sigma}_i$ only. Thus the
masses $\mu_i$ and couplings $\kappa_i$ and $\lambda_i$ are completely independent parameters,
which are not constrained by either $SU(6)$ or $SU(5)$ global symmetries.

The lightest exotic fermion state in the E$_6$CHM must be stable. Indeed, because of baryon number conservation
the low energy effective Lagrangian of the E$_6$CHM is also invariant under the transformations of the baryon triality
\cite{Frigerio:2011zg} which is defined as
\be
\Psi \longrightarrow e^{2\pi i B_3/3} \Psi,\qquad B_3 = (3 B - n_C)_{\mbox{mod}\,\, 3}\,,
\label{10}
\ee
where $B$ is the baryon number of the given multiplet $\Psi$ and $n_C$ is the number of colour indices ($n_C=1$ for the
colour triplet and $n_C=-1$ for ${\bf \overline{3}}$). All SM particles have $B_3=0$, while exotic states carry
either $B_3=1$ or $B_3=2$. Because of this, the lightest exotic state can not decay into SM particles.
If the lightest exotic states were colour triplets or charged fermions then they would be produced during the Big Bang.
These states would survive annihilation and get confined in nuclear isotopes. Different experiments set limits on
the relative concentrations of such stable relics from $10^{-15}$ to $10^{-30}$ per nucleon \cite{Hemmick:1989ns}.
On the other hand theoretical estimates show that if such particles existed in nature their concentration would be much higher
than $10^{-15}$ per nucleon \cite{43}. Thus the models with stable charged exotic particles are basically ruled out.
Moreover, the coupling of the neutral Dirac fermion, which is absolutely stable, to the $Z$--boson should be extremely
suppressed. Otherwise such particles would scatter on nuclei leading to unacceptably large spin--independent cross sections
(for a recent analysis see \cite{Buckley:2013sca})\footnote{Here we assume that the lightest exotic state accounts for all, 
or at least a substantial part, of the observed dark matter density.}. As a consequence, only a Dirac fermion that involves mostly $\eta$ and
$\bar{\eta}$, can be the lightest exotic state in the E$_6$CHM\footnote{It is expected that the relic abundance of the lightest exotic
fermions in the E$_6$CHM is induced by the same mechanism that gives rise to the baryon asymmetry of the Universe.} .

The last three terms in Eq.~(\ref{7}) are induced by the composite partners of the SM fermions. Therefore $\Lambda_1$,
$\Lambda_2$ and $\Lambda_3$ are expected to be of the order of scale $f$. Since naturalness requires that
$\mu_D \sim \mu_Q \sim \mu_L \sim \mu_E \sim f$, whereas the mass of the SM singlet pNGB state $A$ ($m_A$) is expected
to be much smaller than $f$, we restrict our consideration here to the case when all exotic fermions except the lightest one are
heavier than $m_A/2$. In this part of the parameter space the on-shell decays of $A$ into the corresponding exotic fermions are
not kinematically allowed. Integrating out the heavy exotic states, which appear in the usual triangle loop diagrams, one gets
the effective Lagrangian that describes the interactions of pseudoscalar $A$  with the top quark, SM gauge bosons and
dark matter particles (see also \cite{Franceschini:2015kwy}--\cite{Low:2015qep})
\begin{equation}
\begin{array}{c}
\mathcal{L}^A_{eff} = c_1 A B_{\mu\nu} \widetilde{B}^{\mu\nu} + c_2 A W^a_{\mu\nu} \widetilde{W}^{a\mu\nu} +
c_3 A G^{\sigma}_{\mu\nu} \widetilde{G}^{\sigma\mu\nu}\\[2mm]
+  \ds\frac{y_t}{\Lambda_t} A (i\bar{t}_L H t_R + h.c.) + i\lambda_{\eta} A (\bar{\eta}\eta+h.c.)\,,
\end{array}
\label{11}
\end{equation}
where
\begin{equation}
\begin{array}{rcl}
c_1&=&\ds\frac{\alpha_Y}{16\pi}\Biggl[\frac{2\kappa_{D}}{3\mu_{D}} B(x_{D}) + \frac{\kappa_{Q}}{3\mu_{Q}} B(x_{Q})
+  \frac{\lambda_{L}}{\mu_{L}} B(x_{L}) + 2 \frac{\lambda_{E}}{\mu_{E}} B(x_{E}) \Biggr] + \ds\frac{\alpha_Y}{16\pi \Lambda_1}\,,\\[4mm]
c_2&=&\ds\frac{\alpha_2}{16\pi}\Biggl[3 \frac{\kappa_{Q}}{\mu_{Q}} B(x_{Q}) +  \frac{\lambda_{L}}{\mu_{L}} B(x_{L}) \Biggr] +
\ds\frac{\alpha_2}{16\pi \Lambda_2}\,,\\[4mm]
c_3&=&\ds\frac{\alpha_3}{16\pi}\Biggl[\frac{\kappa_{D}}{\mu_{D}} B(x_{D}) + 2 \frac{\kappa_{Q}}{\mu_{Q}} B(x_{Q})\Biggr] +
\ds\frac{\alpha_3}{16\pi \Lambda_3}\,,\\[4mm]
B(x) &=& 2 x \arcsin^2[1/\sqrt{x}]\,,\qquad \mbox{for} \qquad x\ge 1\,.
\end{array}
\label{12}
\end{equation}
In Eq. (\ref{12}) $x_{D}=4\mu_{D}^2/m_A^2$, $x_{Q}=4\mu_{Q}^2/m_A^2$, $x_{L}=4\mu_{L}^2/m_A^2$ and $x_{E}=4\mu_{E}^2/m_A^2$.
It is also convenient to derive an explicit analytical expression for the coupling of the SM singlet pNGB state $A$ to the electromagnetic field.
Using Eqs.~(\ref{11})--(\ref{12}) one finds
\begin{equation}
\mathcal{L}^{A\gamma\gamma}_{eff} = c_{\gamma} A F_{\mu\nu} \widetilde{F}^{\mu\nu}\,,
\label{13}
\end{equation}
$$
c_{\gamma} = c_1 \cos^2\theta_W + c_2 \sin^2\theta_W \simeq
\ds\frac{\alpha}{16\pi}\Biggl[\frac{2\kappa_{D}}{3\mu_{D}} B(x_{D}) + \frac{10 \kappa_{Q}}{3\mu_{Q}} B(x_{Q})
+ 2 \frac{\lambda_{L}}{\mu_{L}} B(x_{L}) + 2 \frac{\lambda_{E}}{\mu_{E}} B(x_{E}) \Biggr]\,,
$$
where $F_{\mu\nu}$ is a field strength associated with the electromagnetic interaction,
$\widetilde{F}^{\mu\nu}=\frac{1}{2}\epsilon^{\mu\nu\lambda\rho} F_{\lambda\rho}$ and
$\theta_W$ is the weak mixing (Weinberg) angle.

From Eqs.~(\ref{12})-(\ref{13}) it follows that the couplings of the SM singlet pNGB state $A$ to the
SM gauge bosons may be considerably larger than their naive estimates that can be obtained using
last terms in the analytical expressions for $c_i$. In particular, this happens if
$\kappa_{D} \sim \kappa_{Q} \sim \lambda_{L}\sim \lambda_{E}\sim 1$ and exotic fermions are
so light that some of them can be discovered at the LHC in the near future. This is the most attractive
scenario that we are going to explore in the next section. It implies that
$\mu_D \sim \mu_Q \sim \mu_L \sim \mu_E \ll f$. In this limit the last terms in the analytical expressions
for $c_i$ (see Eq.~(\ref{12})) are rather small, because $\Lambda_i \sim f$, and, thereby, can be neglected
in our analysis.

\section{Collider Signatures}

At the LHC the SM singlet pNGB state $A$ is predominantly produced through gluon fusion.
The corresponding production cross section is determined by $|c_3|^2$. Since the partial width $\Gamma(A \to gg)$
of the decays $A\to gg$ is also proportional to $|c_3|^2$ it is convenient to present the LHC production cross section
$\sigma_A$ of the pseudoscalar $A$ in the following form \cite{King:2016wep}, \cite{Franceschini:2015kwy}
\be
\sigma_{A}\simeq \ds\frac{K_{gg} C_{gg}}{m_A s}\, \Gamma(A \to gg)\,,\qquad
\Gamma(A \to gg)=\ds\frac{2 m_A^3}{\pi} |c_3|^2\,,
\label{14}
\ee
where $\sqrt{s}\simeq 13\,\mbox{TeV}$. To simplify our analysis we assume that $m_A=750\,\mbox{GeV}$
so that in our calculations we can use the same values of the dimensionless partonic integral $C_{gg}$ and K factor $K_{gg}$
as in \cite{Franceschini:2015kwy}, i.e. $C_{gg}=2137$ and $K_{gg}=1.48$. In this case the LHC production cross section
can be approximately estimated as
\be
\sigma_{A}\simeq 7.3\, \mbox{fb}\times \left( \ds\frac{\Gamma(A \to gg)}{m_{A}}\times 10^{6}\right)\,.
\label{15}
\ee

Although as mentioned before we restrict our consideration here to the scenario when all charged exotic fermions
are heavier than $m_A/2$ we assume that the mass of the lightest exotic fermion is quite close to $m_A/2$ and the Yukawa
coupling $\lambda_{\eta}\gtrsim 1$. This gives rise to sufficiently large annihilation cross section for
$\eta\, \bar{\eta} \to \mbox{SM particles}$ and therefore should enable the lightest exotic fermion state to account
for all or some of the observed cold dark matter density. Thus we allow the lightest exotic fermion state to be either lighter
or heavier than $m_A/2$. If $\mu_{\eta} > m_A/2$ then the total decay width $\Gamma_A$ tends to be sufficiently small.
When $\mu_{\eta} < m_A/2$ the tree-level decays of $A$ into $\eta\, \bar{\eta}$ are kinematically allowed so that $\Gamma_A$
can be rather large. Here we examine these two possibilities separately.

%In Eq.~(\ref{16}) the partial decay width associated with the decays of the SM singlet pNGB state into a pair of photons
%is given by
%\be
%\Gamma(A \to \gamma\gamma)=\ds\frac{m_A^3}{4 \pi} |c_{\gamma}|^2\,.
%\label{17}
%\ee

\subsection{Narrow width case}

Let us first consider the scenario with $\mu_{\eta} > m_A/2$.  The pseudoscalar $A$ can decay into fermion-antifermion pairs.
In particular, if $m_A$ is larger than $2 m_t$, where $m_t$ is a top quark mass, the partial decay width, that corresponds to the
decay mode $A \to t\bar{t}$, is given by
\be
\Gamma(A \to t\bar{t})=\ds\frac{3 m_A m_t^2}{8\pi \Lambda_t^2}\sqrt{1-\ds\frac{4 m_t^2}{m_A^2}}\,.
\label{18}
\ee
The partial decay width (\ref{18}) decreases when $\Lambda_t$ and $f$ increase. For $\Lambda_t \gg 10\,\mbox{TeV}$
the value of $\Gamma(A \to t\bar{t})$ becomes rather small. Nevertheless this region of the parameter space is strongly
disfavoured by naturalness arguments. Indeed, in this case an extremely large fine--tuning is needed to obtain the EW scale $v\ll f$.
In our numerical analysis we set $f\simeq 10\,\mbox{TeV}$ and study two different scenarios with $\Lambda_t=\sqrt{15} f\gg 10\,\mbox{TeV}$
and $\Lambda_t=\sqrt{\frac{60}{49}} f\simeq 1.1 f\sim 10\,\mbox{TeV}$ associated with $t^c$ being predominantly
the appropriate components of ${\bf 20}^t$ and ${\bf 15}^t$ of $SU(6)$ respectively. For so large values of  $\Lambda_t$
the decay rates $A \to  b \bar{b}$ and $A\to \tau\bar{\tau}$ become negligibly small and can be ignored in the leading approximation.

The analytical expressions for the partial widths associated with the decays $A\to WW, ZZ, \gamma\gamma$ and $\gamma Z$ can be
presented in the following form
\begin{equation}
\Gamma(A \to W W)=\ds\frac{m_A^3}{2 \pi} |c_2|^2 \left(1-\ds\frac{4 M_W^2}{m_A^2}\right)^{3/2},\qquad\qquad\qquad\qquad
\label{19}
\end{equation}
\begin{equation}
\Gamma(A \to ZZ)=\ds\frac{m_A^3}{4 \pi} \Biggl|c_1 \sin^2 \theta_W + c_2 \cos^2 \theta_W \Biggr|^2 \left(1-\ds\frac{4 M_Z^2}{m_A^2}\right)^{3/2}\,,\\[3mm]
\label{20}
\end{equation}
\begin{equation}
\Gamma(A \to \gamma\gamma)=\ds\frac{m_A^3}{4 \pi} |c_{\gamma}|^2\,,\qquad\qquad\qquad\qquad\qquad\qquad\qquad\qquad
\label{17}
\end{equation}
\begin{equation}
\Gamma(A \to \gamma Z)= \ds\frac{m_A^3}{8 \pi} \sin^2 2\theta_W |c_1-c_2|^2 \left(1-\ds\frac{M_Z^2}{m_A^2}\right)^3\,.\qquad\qquad
\label{21}
\end{equation}
If $\mu_D \sim \mu_Q \sim \mu_L \sim \mu_E$ then from Eqs.~(12) it follows that $|c_2|$ is substantially larger than
$|c_1|$. This happens because $c_2$ and $c_1$ are proportional to the $SU(2)_W$ and $U(1)_Y$ gauge couplings respectively and
at low energies $\alpha_2$ is considerably bigger than $\alpha_Y$. When $|c_2|$ is considerably larger than $|c_1|$ the partial decay
widths $\Gamma(A \to W W)$ and $\Gamma(A \to ZZ)$ should be substantially bigger as compared with $\Gamma(A \to \gamma\gamma)$
and $\Gamma(A \to \gamma Z)$ because the value of $\sin^2 \theta_W$ is quite small. On the other hand in this case $|c_2|\ll |c_3|$
since $\alpha_2$ is much smaller than $\alpha_3$. Thus one can expect that in general  $\Gamma(A \to gg)$ is substantially larger
than the partial decay widths (\ref{19})--(\ref{21}).

\begin{figure}
\hspace*{0cm}{$\mbox{BR}(A \to t\bar{t}, gg, WW, ZZ, \gamma\gamma, \gamma Z)$}
\hspace*{2.5cm}{$\mbox{BR}(A \to t\bar{t}, gg, WW, ZZ, \gamma\gamma, \gamma Z)$}\\[-4mm]
\includegraphics[width=75mm,height=55mm]{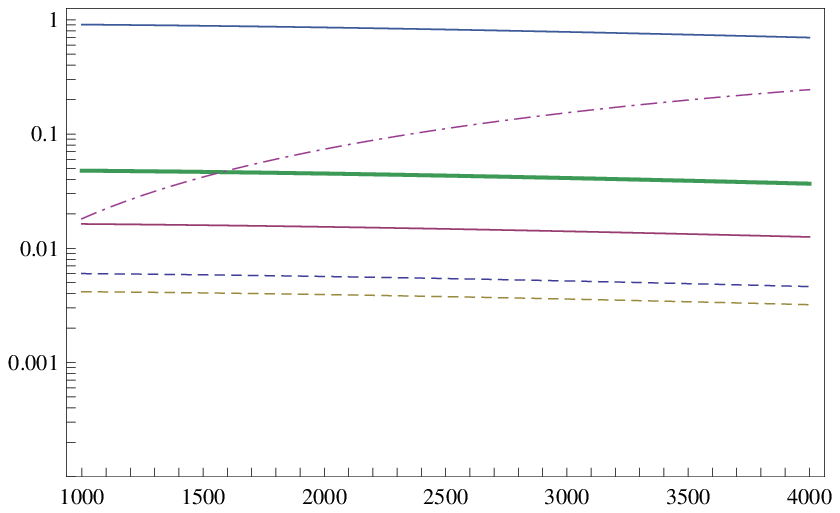}\qquad
\includegraphics[width=75mm,height=55mm]{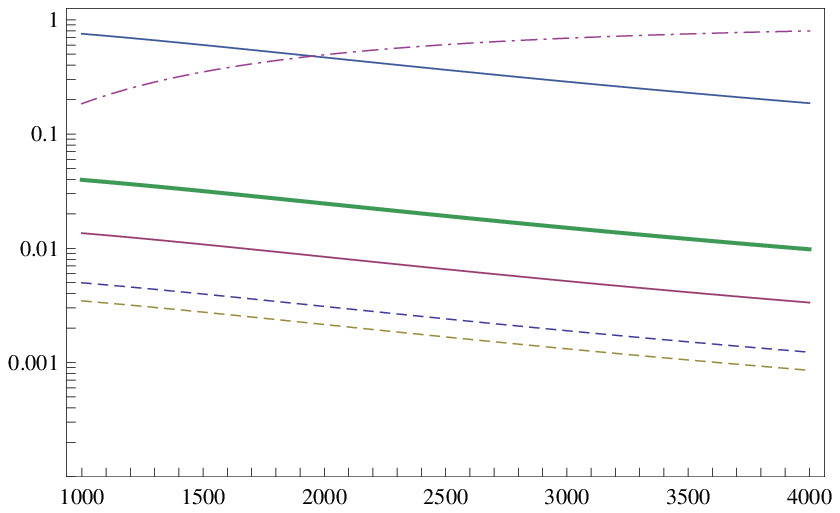}\\[-5mm]
\hspace*{3.5cm}{$\mu_0$}\hspace*{8.5cm}{$\mu_0$}\\[1mm]
\hspace*{3.5cm}{\bf (a)}\hspace*{8cm}{\bf (b) }\\[4mm]
\hspace*{0cm}{$\mbox{BR}(A \to t\bar{t}, gg, WW, ZZ, \gamma\gamma, \gamma Z)$}
\hspace*{2.5cm}{$\mbox{BR}(A \to t\bar{t}, gg, WW, ZZ, \gamma\gamma, \gamma Z)$}\\[-4mm]
\includegraphics[width=75mm,height=55mm]{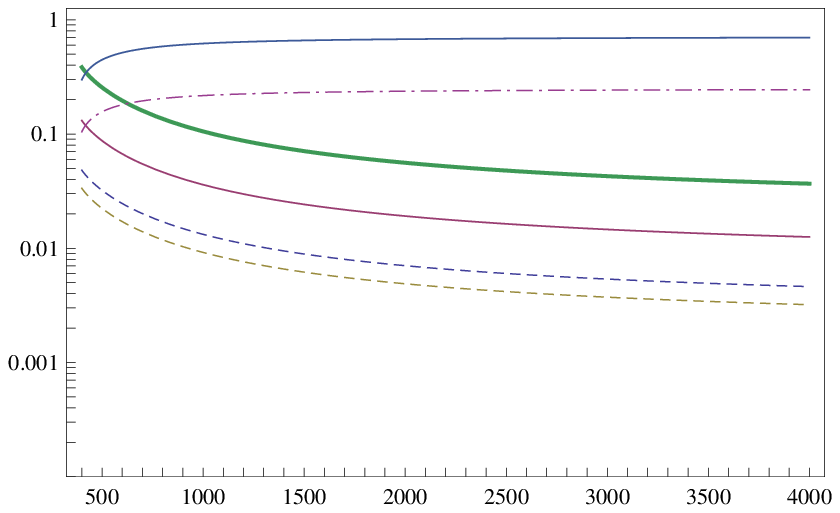}\qquad
\includegraphics[width=75mm,height=55mm]{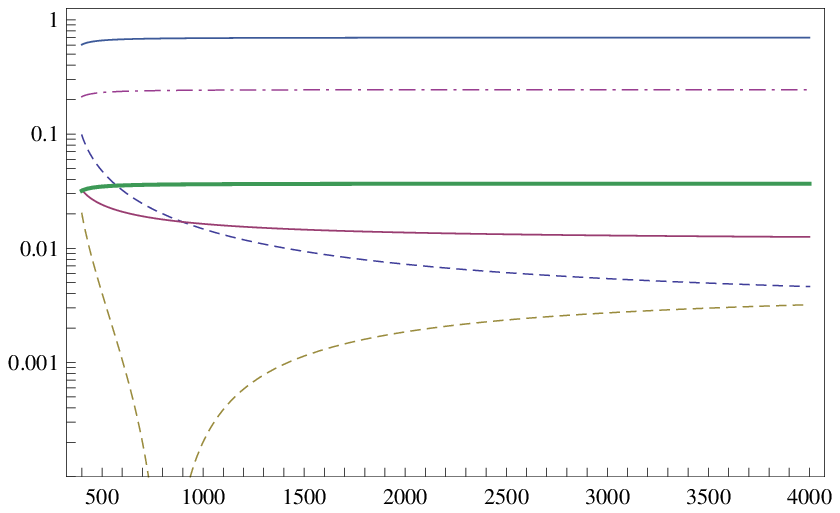}\\[-5mm]
\hspace*{3.5cm}{$\mu_L$}\hspace*{8.5cm}{$\mu_E$}\\[1mm]
\hspace*{3.5cm}{\bf (c)}\hspace*{8cm}{\bf (d) }\\[2mm]
\caption{The branching ratios of the decays of the pNGB state $A$ into $t\bar{t}$ (dashed--dotted lines), $gg$ (highest solid lines),
$\gamma\gamma$ (highest dashed lines), $WW$ (thick solid lines), $ZZ$ (lowest solid lines) and $\gamma Z$ (lowest dashed lines)
for $m_A=750\, \mbox{GeV}$, $\kappa_D = \kappa_Q = \lambda_L = \lambda_E = \sigma = 1.5$ and $f=10\,\mbox{TeV}$.
In {\it (a)} the branching fractions of the decays of $A$ are given as a function of $\mu_Q=\mu_D=\mu_L=\mu_E=\mu_0$ for
$\Lambda_t=\sqrt{15} f$. In {\it (b)} the branching ratios of the decays of $A$ are shown as a function of
$\mu_Q=\mu_D=\mu_L=\mu_E=\mu_0$ for $\Lambda_t=\sqrt{\frac{60}{49}} f$. In {\it (c)} the branching fractions of the decays
of $A$ are presented as a function of $\mu_L$ for $\mu_Q=\mu_D=\mu_E=4\,\mbox{TeV}$ and $\Lambda_t=\sqrt{15} f$.
In {\it (d)} the branching ratios of the decays of $A$ are shown as a function of $\mu_E$ for $\mu_Q=\mu_D=\mu_L=4\,\mbox{TeV}$
and $\Lambda_t=\sqrt{15} f$.}
\label{fig1}
\end{figure}

Our results are summarised in Fig.~1-3. In Fig.~1 the dependence of the branching ratios of the SM singlet pNGB state $A$ on
the masses of exotic fermions and $\Lambda_t$ is examined. The branching ratios of the pseudoscalar $A$ are given by
\begin{equation}
\begin{array}{ll}
\mbox{BR}(A \to gg)=\ds\frac{\Gamma(A \to gg)}{\Gamma_{A}}\,, \qquad &
\mbox{BR}(A \to \gamma\gamma)=\ds\frac{\Gamma(A \to \gamma\gamma)}{\Gamma_{A}}\,,\\[3mm]
\mbox{BR}(A \to t\bar{t})=\ds\frac{\Gamma(A \to t\bar{t})}{\Gamma_{A}}\,, \qquad &
\mbox{BR}(A \to Z\gamma)=\ds\frac{\Gamma(A \to Z\gamma)}{\Gamma_{A}}\,,\\[3mm]
\mbox{BR}(A \to Z Z)=\ds\frac{\Gamma(A \to ZZ)}{\Gamma_{A}}\,,\qquad &
\mbox{BR}(A \to W W)=\ds\frac{\Gamma(A \to W W)}{\Gamma_{A}}\,,
\end{array}
\label{22}
\end{equation}
where $\Gamma_{A}$ is a total decay width of $A$.
To simplify our analysis here we focus on the scenarios with $\kappa_D = \kappa_Q = \lambda_L = \lambda_E = \sigma$.
In Figs.~1a and 1b the masses of all exotic fermions are set to be equal, i.e. $\mu_D = \mu_Q = \mu_L = \mu_E= \mu_0$,
whereas $\Lambda_t=\sqrt{15} f$ and $\Lambda_t=\sqrt{\frac{60}{49}} f$ respectively. The absence of new particles
carrying colour with masses below $1\,\mbox{TeV}$, which should have sufficiently large LHC production cross section,
implies that all exotic, coloured fermions and scalar coloured triplet in the E$_6$CHM have to be quite heavy.
At the same time we choose $\mu_0$ to be substantially lower than $10\,\mbox{TeV}$ to ensure that the production cross
section of the SM singlet pNGB state $A$ is large and exotic quarks may be observed at the LHC in the near future.

In the composite Higgs models with $f\sim 1\,\mbox{TeV}$ the pNGB pseudoscalar state $A$ tends to decay mainly into
$t\bar{t}$ \cite{Franceschini:2015kwy}--\cite{Low:2015qep}. In the model under consideration such low values of the scale
$f$ are ruled out. As follows from Fig.~1a and 1b large values of $\Lambda_t$ associated with $f\simeq 10\,\mbox{TeV}$
lead to such a suppression of the partial decay width (\ref{18}) that $\Gamma(A \to t\bar{t})$ becomes comparable with
the one--loop induced decay rate $A \to gg$ if $\sigma\gtrsim 1$. Moreover one can see that for $\sigma=1.5$ and
relatively low values of $\mu_0\lesssim 2\,\mbox{TeV}$ the branching ratio of $A\to t\bar{t}$ is not the dominant one
even when $\Lambda_t \sim f$. $\mbox{BR}(A \to t\bar{t})$ grows with increasing $\mu_0$. Nevertheless Fig.~1a
demonstrates that for $\sigma=1.5$ and very large $\Lambda_t\gg f$ (i.e. $\Lambda_t\simeq \sqrt{15} f$)
$A \to gg$ remains the main decay mode of the pseudoscalar $A$. Whilst $A \to gg$ and $A \to t\bar{t}$ are
two main decay channels of the SM singlet state $A$, the branching ratios of $A\to WW$ and $A\to ZZ$ tend to be
the third and the fourth largest ones. As was discussed before, $\mbox{BR}(A \to \gamma\gamma)$ and
$\mbox{BR}(A \to \gamma Z)$ are considerably smaller and vary between $10^{-2}$ and $10^{-3}$.
The branching ratios of $A\to b\bar{b}$ and $A\to \tau\bar{\tau}$, which are not shown in Fig.~1,
are always less than $10^{-3}$ and $10^{-4}$ respectively.

The hierarchical structure of the coefficients in the effective Lagrangian (\ref{11}) $|c_1|\ll |c_2|\ll |c_3|$ is not
always valid. Fig.~1c and 1d indicate that there may be some special cases, when it gets spoiled, if the masses
of exotic fermions, for example, are very different. In order to make $c_2$ large it is enough to assume that
$\mu_L\ll \mu_D = \mu_Q = \mu_E$. In Fig.~1c the variations of the branching fractions of the SM singlet
pNGB state $A$ is studied as a function of $\mu_L$ for $\sigma=1.5$ and $\mu_D = \mu_Q = \mu_E = 4\,\mbox{TeV}$.
The coefficient $c_2$ as well as branching ratios of $A\to WW, ZZ, \gamma\gamma$ and $\gamma Z$ grow with
decreasing $\mu_L$. For $\mu_L\simeq 500\,\mbox{GeV}$ the coefficient $c_3$ is considerably smaller than $c_2$.
As a result the branching ratio of $A\to WW$ becomes comparable with $\mbox{BR}(A \to gg)$ and is substantially
larger than $\mbox{BR}(A \to t\bar{t})$. In this case $\mbox{BR}(A \to t\bar{t})\simeq \mbox{BR}(A \to ZZ)$
whereas $\mbox{BR}(A \to \gamma\gamma)$ and $\mbox{BR}(A \to \gamma Z)$ are larger than $10^{-2}$
but still less than other branching ratios of the decays $A\to gg, WW, t\bar{t}$ and $ZZ$. The coefficient $c_2$
and $c_1$ in the effective Lagrangian (\ref{11})  diminish with increasing $\mu_L$ together with
$\mbox{BR}(A \to WW, ZZ, \gamma\gamma, \gamma Z)$ and for $\mu_L\gtrsim 1.5\,\mbox{TeV}$
the qualitative pattern of branching ratios is almost the same as the one presented in Fig.~1a
when $\mu_0\gtrsim 2\,\mbox{TeV}$.

\begin{figure}
\hspace*{0cm}{$\frac{\Gamma_A}{m_A} \times 10^5$}
\hspace*{6.5cm}{$\sigma_A[\mbox{pb}]$}\\[-4mm]
\includegraphics[width=75mm,height=55mm]{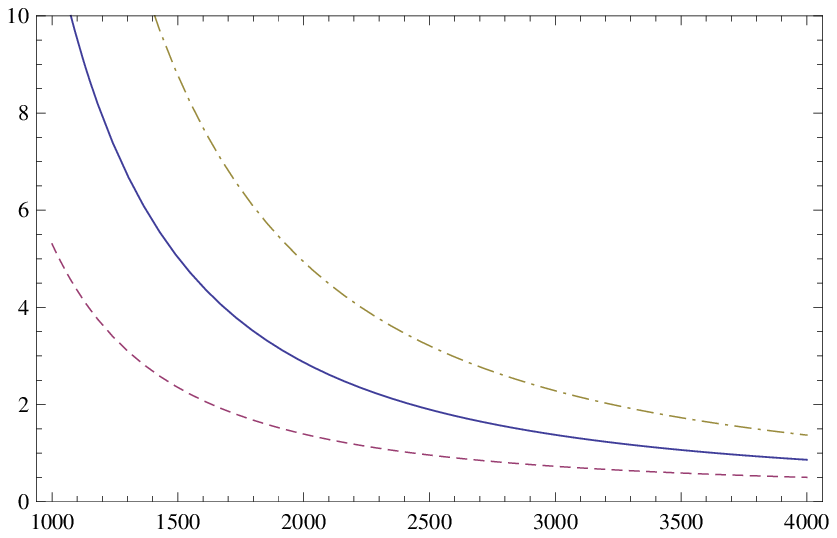}\qquad
\includegraphics[width=75mm,height=55mm]{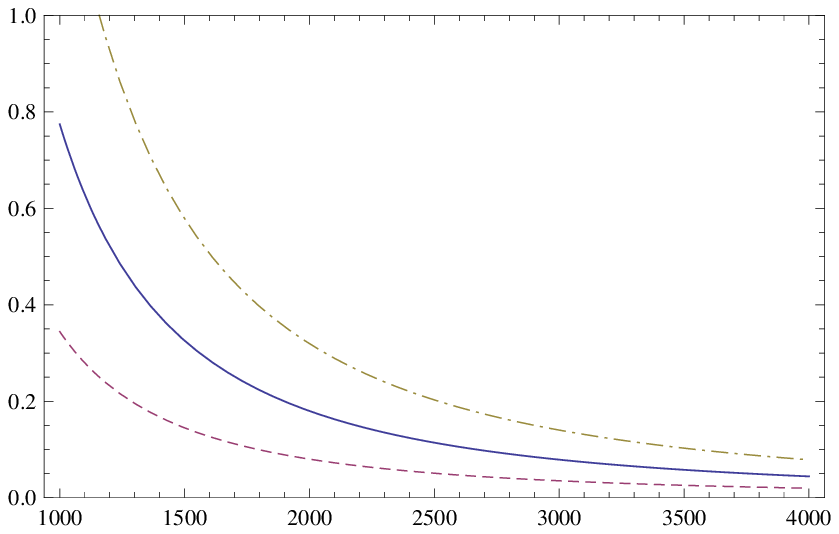}\\[-5mm]
\hspace*{3.5cm}{$\mu_0$}\hspace*{8.5cm}{$\mu_0$}\\[1mm]
\hspace*{3.5cm}{\bf (a)}\hspace*{8cm}{\bf (b) }\\[4mm]
\hspace*{0cm}{$\frac{\Gamma_A}{m_A} \times 10^5$}
\hspace*{6.5cm}{$\sigma_A[\mbox{fb}]$}\\[-4mm]
\includegraphics[width=75mm,height=55mm]{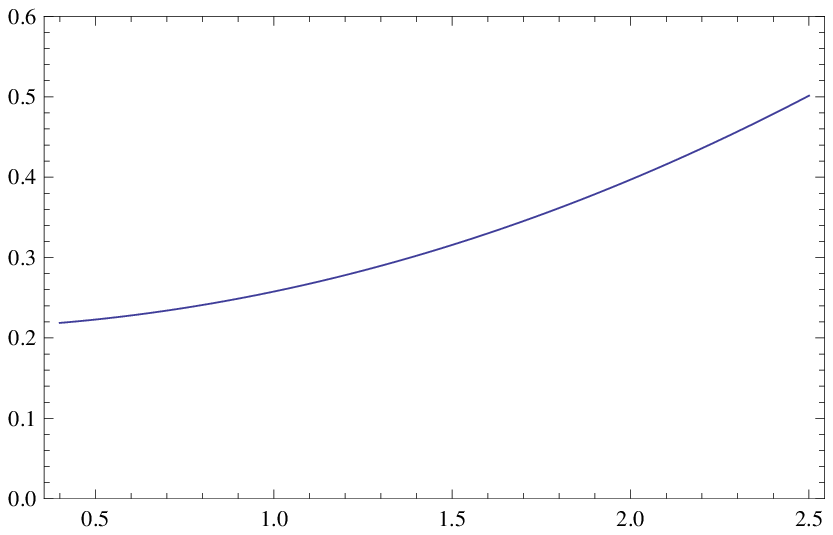}\qquad
\includegraphics[width=75mm,height=55mm]{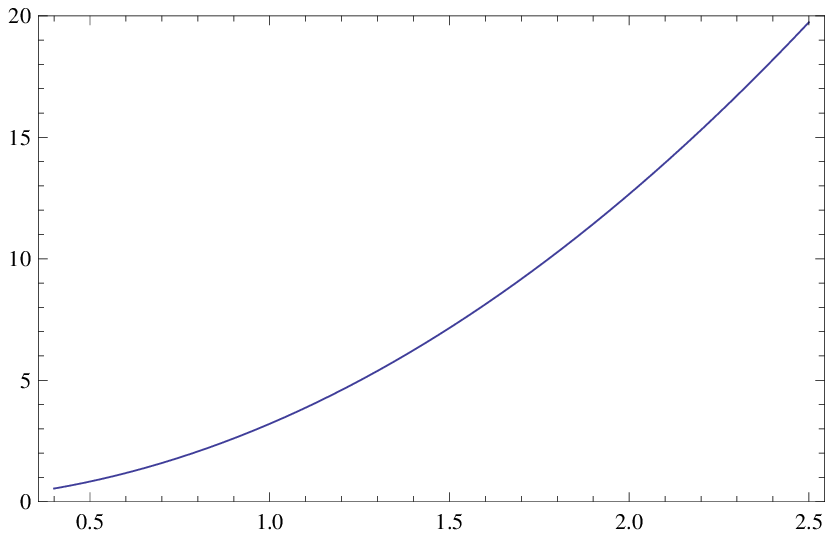}\\[-5mm]
\hspace*{3.5cm}{$\sigma f/\mu_0$}\hspace*{7cm}{$\sigma f/\mu_0$}\\[1mm]
\hspace*{3.5cm}{\bf (c)}\hspace*{8cm}{\bf (d) }\\[2mm]
\caption{The ratio $\Gamma_A/m_A$ and the LHC production cross section $\sigma_A$ of the pNGB state $A$ are shown for
$\mu_Q=\mu_D=\mu_L=\mu_E=\mu_0$, $\kappa_D = \kappa_Q = \lambda_L = \lambda_E = \sigma$, $\Lambda_t=\sqrt{15} f$,
$f=10\,\mbox{TeV}$ and $m_A=750\,\mbox{GeV}$. In {\it (a)} the ratio $\Gamma_A/m_A$ is presented as a function of $\mu_0$ for
$\sigma=2$ (dashed--dotted line), $\sigma=1.5$ (solid line) and $\sigma=1$ (dashed line). In {\it (b)} the cross section $\sigma_A$
is presented as a function of $\mu_0$ for $\sigma=2$ (dashed--dotted line), $\sigma=1.5$ (solid line) and $\sigma=1$ (dashed line).
In {\it (c)} we show the ratio $\Gamma_A/m_A$ as a function of $\sigma f/\mu_0$ for $\mu_0=4\,\mbox{TeV}$.  In {\it (d)} the LHC
production cross section $\sigma_A$ is shown as a function of $\sigma f/\mu_0$ for $\mu_0=4\,\mbox{TeV}$.}
\label{fig2}
\end{figure}

The coefficient $c_1$ may become comparable with $c_3$ and much larger than $c_2$ if $\mu_E\ll \mu_D = \mu_Q = \mu_L$.
This is illustrated in Fig.~1d, where we plot the branching fractions of the decays of the pseudoscalar $A$ as a function of
$\mu_E$ for $\sigma=1.5$ and $\mu_D = \mu_Q = \mu_L = 4\,\mbox{TeV}$. Whilst $c_2$ and $c_3$ do not change when
$\mu_E$ varies the coefficients $c_1$ and $c_{\gamma}$ increase with decreasing $\mu_E$. For $\mu_E\simeq 500\,\mbox{GeV}$
the coefficients $c_1$ and $c_{\gamma}$ are much bigger than $c_2$ and $c_3 \lesssim c_1, c_{\gamma}$.
As we see from Fig.~1d, when $|c_2|\ll |c_1|$ the partial decay widths $\Gamma(A\to WW)$, $\Gamma(A \to ZZ)$ and $\Gamma(A \to \gamma Z)$
are considerably smaller than $\Gamma(A \to \gamma\gamma)$. In this case
$A\to gg$ and $t\bar{t}$ are the dominant decay modes of this state whereas $\mbox{BR}(A \to \gamma\gamma)$ may be
the third largest branching ratio if $\mu_E$ is sufficiently small. As follows from Fig.~1d $\mbox{BR}(A \to \gamma\gamma)$
can be as large as $10\%$ and diminishes rapidly with increasing $\mu_E$. When $\mu_E\gtrsim 1.5\,\mbox{TeV}$
the branching ratio of $A \to \gamma\gamma$ is the fifth largest one and considerably smaller than $\mbox{BR}(A \to WW)$
and $\mbox{BR}(A \to ZZ)$ which are the third and fourth largest branching fractions respectively.

In Figs.~2 we consider the dependence of the total decay width $\Gamma_A$ of the SM singlet pNGB state $A$ and its LHC production
cross section $\sigma_A$ on $\sigma$ and $\mu_0$. In particular, in Fig.~2a and 2b the variations of $\Gamma_A/m_A$ and $\sigma_A$
are explored as a function of $\mu_0$ in the case of the scenario with $\sigma=1.5$ which was used before to demonstrate the
dependence of branching ratios on the exotic fermion masses (see Fig.~1). One can see that the total decay width of the pseudoscalar $A$
and its LHC production cross section decrease very rapidly with increasing $\mu_{0}$ and grow if $\sigma$ increases.
When $\mu_0$ changes from $1\,\mbox{TeV}$ to $4\,\mbox{TeV}$ the ratio $\Gamma_A/m_A$ diminishes from $10^{-4}$ to $10^{-5}$
whereas $\sigma_A$ decreases from $1\,\mbox{pb}$ to $50\,\mbox{fb}$.

\begin{figure}
\hspace*{0cm}{$\mbox{BR}(A \to t\bar{t}, gg, WW, ZZ, \gamma\gamma, \gamma Z)$}
\hspace*{2.5cm}{$\mbox{BR}(A \to t\bar{t}, gg, WW, ZZ, \gamma\gamma, \gamma Z)$}\\[-4mm]
\includegraphics[width=75mm,height=55mm]{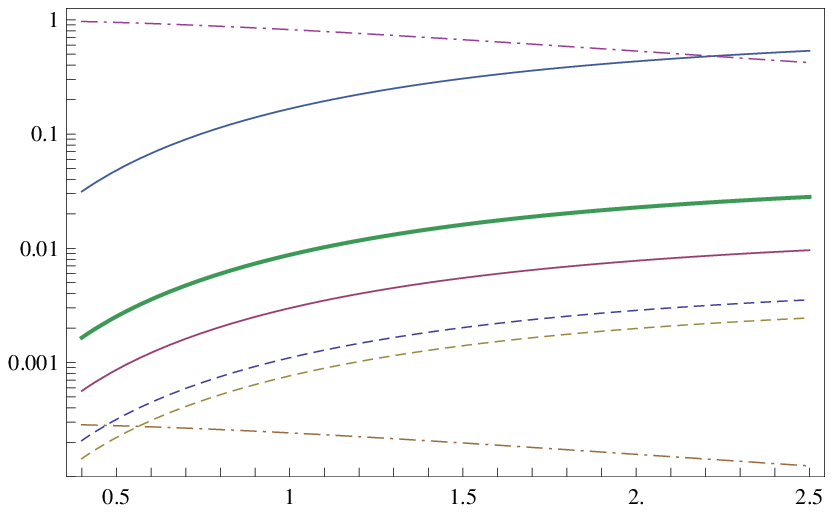}\qquad
\includegraphics[width=75mm,height=55mm]{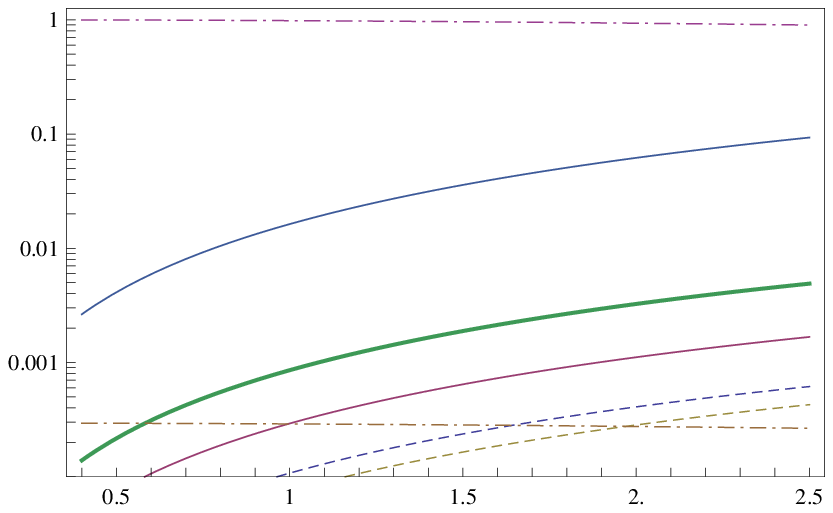}\\[-5mm]
\hspace*{3.5cm}{$\sigma f/\mu_0$}\hspace*{7cm}{$\sigma f/\mu_0$}\\[1mm]
\hspace*{3.5cm}{\bf (a)}\hspace*{8cm}{\bf (b) }\\[2mm]
\caption{The branching ratios of the decays of the pseudoscalar $A$ into $t\bar{t}$ (highest dashed--dotted lines),
$b\bar{b}$ (lowest dashed--dotted lines), $gg$ (highest solid lines), $\gamma\gamma$ (highest dashed lines),
$WW$ (thick solid lines), $ZZ$ (lowest solid lines) and $\gamma Z$ (lowest dashed lines) for $m_A=750\, \mbox{GeV}$,
$\kappa_D = \kappa_Q = \lambda_L = \lambda_E = \sigma$, $f=10\,\mbox{TeV}$ and
$\mu_Q=\mu_D=\mu_L=\mu_E=\mu_0=4\,\mbox{TeV}$. In {\it (a)} the branching fractions of the decays of $A$
are presented as a function of $\sigma f/\mu_0$ for $\Lambda_t=\sqrt{15} f$. In {\it (b)} the branching ratios of the
decays of $A$ are shown as a function of $\sigma f/\mu_0$ for $\Lambda_t=\sqrt{\frac{60}{49}} f$.}
\label{fig3}
\end{figure}

Thus far we have focussed on the scenarios in which all exotic fermions are relatively light, i.e. their masses are considerably lower than the scale $f$,
and their couplings to SM singlet pNGB state $A$ are larger than unity. However such scenarios are excessively fine-tuned. Indeed, the
masses of the exotic fermions tend to be of the order of $\kappa_i f$ and $\lambda_i f$. If $\kappa_i \sim \lambda_i \sim 1$ some fine-tuning
is required to keep exotic fermions substantially lighter than the scale $f$. In this context the scenarios, in which the ratios of
the charged exotic fermion masses to the corresponding Yukawa couplings are of the order of $f$, look more natural. It is also worth
pointing out that in the limit when all charged exotic fermions are much heavier than pseudoscalar $A$ the coefficients $c_i$ in the
effective Lagrangian (\ref{11}) depend only on such ratios. In Fig.~2c and 2d we examine the dependence of $\Gamma_A/m_A$ and
$\sigma_A$ on the ratio $\sigma f/\mu_0$, which is taken to be of order unity, for $\kappa_D = \kappa_Q = \lambda_L = \lambda_E = \sigma$,
$m_A=750\,\mbox{GeV}$ and $f=10\,\mbox{TeV}$. Although we set $\mu_D = \mu_Q = \mu_L = \mu_E= \mu_0=4\,\mbox{TeV}$
the numerical values of $\Gamma_A/m_A$ and $\sigma_A$ should not depend on $\mu_0$ whilst $\mu_0 \gg M_A$ and the ratio $\mu_0/\sigma$
is fixed. One can see that this more natural scenario implies that $\Gamma_A/m_A$ and $\sigma_A$ are substantially lower
than those presented in Fig.~2a and 2b. In particular, when $\sigma f/\mu_0$ is close to unity, the LHC production
cross section of the pseudoscalar $A$ is just a few fb.

The variations of the branching fractions of the decays of the SM singlet pNGB
state $A$ associated with this scenario are studied in Fig.~3a and 3b. In this case the pseudoscalar $A$ decays mainly into $t\bar{t}$.
The branching ratio of the decay $A\to gg$ can be considerably smaller than $\mbox{BR}(A \to t\bar{t})$ but is much larger than
$\mbox{BR}(A \to WW)$ which is the third largest one. As before $\mbox{BR}(A \to ZZ)$ is smaller than $\mbox{BR}(A \to WW)$
but is significantly larger than $\mbox{BR}(A \to \gamma\gamma)$ whereas the branching ratio of the decay $A\to \gamma\gamma$ is
bigger than $\mbox{BR}(A \to \gamma Z)$. As follows from Fig.~3a and 3b the branching ratio of the decay $A\to b\bar{b}$ may be
comparable with $\mbox{BR}(A \to ZZ)$, $\mbox{BR}(A \to \gamma\gamma)$, $\mbox{BR}(A \to \gamma Z)$ and even
$\mbox{BR}(A \to WW)$.

In our analysis we explore the production and decays of the pseudoscalar $A$ in the part of the E$_6$CHM parameter space
where $\sigma_A$ is considerably larger as compared with the LHC production cross sections of the SM singlet pNGB states in
the simplest composite Higgs models with $f\gtrsim 10\,\mbox{TeV}$. Such enhancement of $\sigma_A$ can be caused by the
presence of exotic quarks in the E$_6$CHM only if the contributions to $c_3$ from different exotic quark multiplets as well as composite
partners of the SM particles do not cancel each other. If the cancellation of different contributions to $c_3$ takes place then $\sigma_A$
may be substantially smaller than 1 fb. Indeed, in this case the partial width of the decay $A\to gg$ can be estimated as
\be
\Gamma(A \to gg)=\ds\frac{2 m_A^3}{\pi f^2} \left(\ds\frac{\alpha_3 }{16\pi} \right)^2 \times a^2\,,
\label{221}
\ee
where $a=16\pi f c_3/\alpha_3$ is a dimensionless parameter. Large cancellation of different contributions to $c_3$ corresponds to $a\ll 1$.
Then for $m_A\simeq 750\,\mbox{GeV}$ using Eq.~(\ref{15}) one can obtain
\be
\sigma_{A}\simeq 0.086\, \mbox{fb}\times a^2\,.
\label{222}
\ee
Naturalness requires the parameter $a$ to be of the order of unity. Varying $a$
around unity one finds that in the case of reasonably natural scenarios
the cross section $\sigma_A$ changes from $0.01\,\mbox{fb}$ to a few fb.
This relatively small production cross section of the SM singlet pNGB state $A$
makes its observation at the LHC rather problematic.

%The results of our numerical analysis indicates that in this part of the parameter space the
%pseudoscalar $A$ still mainly decays into a pair of gluons. These in turn are very difficult to detect at the LHC if the corresponding
%cross section is sufficiently small.

% On the other hand, the decays of the pseudoscalar $A$ into
% a pair of gluons are quite difficult to detect since the total LHC production cross section of this state remains rather small.

%Although the branching ratios of $A \to WW$ and $A \to ZZ$ can be substantially larger than the
%branching ratio of $A\to \gamma\gamma$, their experimental detection is more problematic because the $W$ and $Z$ decay mainly into quarks.

\subsection{Large width scenario }

\begin{figure}
\hspace*{0cm}{$\Gamma(A\to \eta\bar{\eta})/m_A$}
\hspace*{5.5cm}{$\mbox{BR}(A \to t\bar{t}, gg, WW, ZZ, \gamma\gamma, \gamma Z)\times 10^3$}\\[-4mm]
\includegraphics[width=75mm,height=55mm]{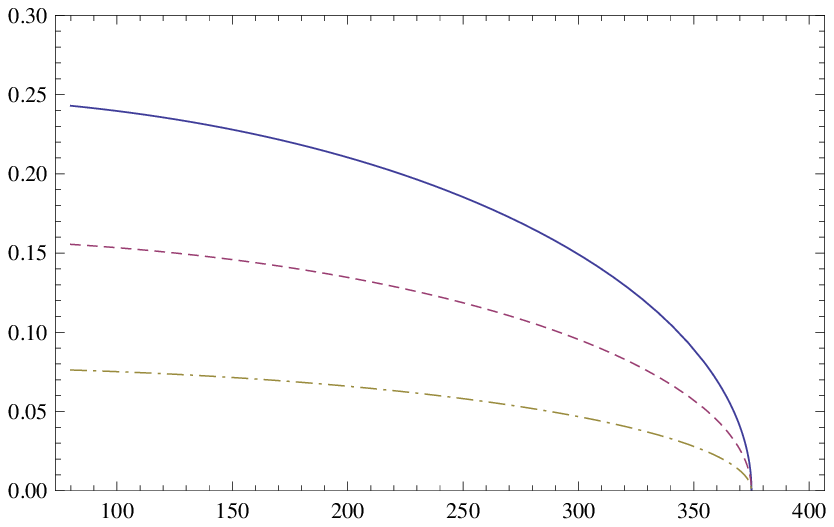}\qquad
\includegraphics[width=75mm,height=55mm]{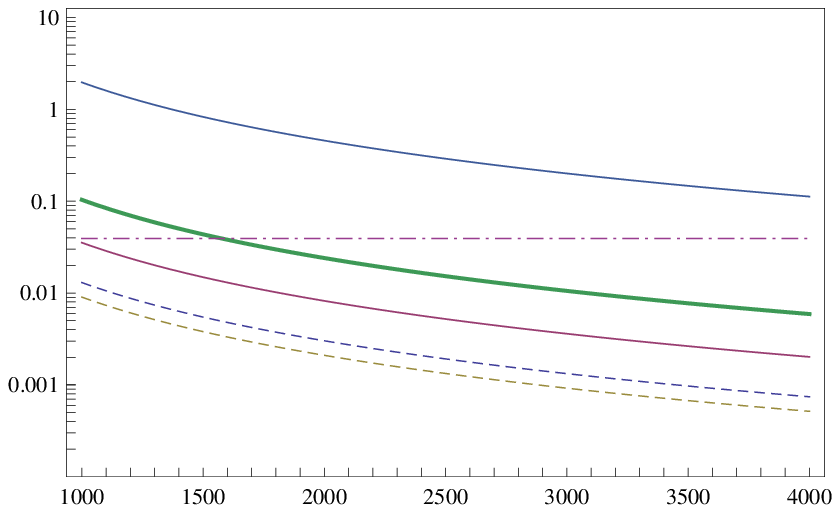}\\[-5mm]
\hspace*{3.5cm}{$\mu_{\eta}$}\hspace*{8cm}{$\mu_0$}\\[1mm]
\hspace*{3.5cm}{\bf (a)}\hspace*{8cm}{\bf (b) }\\[2mm]
\caption{The ratio $\Gamma(A\to \eta\bar{\eta})/m_A$ and the branching ratios of the decays of the pNGB state $A$ are
presented for $m_A=750\,\mbox{GeV}$. In {\it (a)} the ratio $\Gamma(A\to \eta\bar{\eta})/m_A$
is shown as a function of the mass of the lightest exotic fermion $\mu_{\eta}$ for $\lambda_{\eta}=2.5$ (solid line),
$\lambda_{\eta}=2$ (dashed line) and $\lambda_{\eta}=1.4$ (dashed--dotted line).
In {\it (b)} the branching ratios of the decays of $A$ into $t\bar{t}$ (dashed--dotted line), $gg$ (highest solid line),
$\gamma\gamma$ (highest dashed line), $WW$ (thick solid line), $ZZ$ (lowest solid line) and $\gamma Z$ (lowest dashed line)
are given as a function of $\mu_Q=\mu_D=\mu_L=\mu_E=\mu_0$ for $\kappa_D = \kappa_Q = \lambda_L = \lambda_E=\lambda_{\eta}= 1.5$,
$\Lambda_t=\sqrt{15} f$, $f=10\,\mbox{TeV}$ and $\mu_{\eta}=300\,\mbox{GeV}$.}
\label{fig4}
\end{figure}

Let us now explore the scenario with $\mu_{\eta}\lesssim m_A/2$, so that the decays of $A$ into $\eta\, \bar{\eta}$ are
kinematically allowed. Fig.~4a indicates that the partial width associated with the decays $A\to \eta\, \bar{\eta}$ tends to be rather
large in this case if the Yukawa coupling $\lambda_{\eta}$ is larger than unity. As a consequence the total width $\Gamma_A$
of the pseudoscalar $A$ is also large and $\Gamma_A \simeq \Gamma(A\to \eta\bar{\eta})$. Here again we assume that
the mass of the lightest exotic fermion is rather close to $m_A/2$ resulting in sufficiently big cross section for
$\eta\, \bar{\eta} \to \mbox{SM particles}$, and set $\mu_D = \mu_Q = \mu_L = \mu_E= \mu_0$ as well as
$\kappa_D = \kappa_Q = \lambda_L = \lambda_E = \sigma$.

As one can see from Fig.~4b the large total width $\Gamma_A$
leads to the strong suppression of all branching fractions of the decays of $A$ except $\mbox{BR}(A \to \eta\bar{\eta})$
as compared with the narrow width scenarios. Fig.~4b demonstrates that for $\sigma=\lambda_{\eta}= 1.5$,
$\Lambda_t=\sqrt{15} f$, $f=10\,\mbox{TeV}$ and $\mu_{\eta}=300\,\mbox{GeV}$ the branching ratio of
$A\to gg$ is the second largest when $\mu_0$ varies from $1\,\mbox{TeV}$ to $4\,\mbox{TeV}$.
The branching fractions of $A\to WW$ and $A\to t\bar{t}$ can be either the third or fourth largest. If $\Lambda_t$ is smaller,
say around $\sqrt{\frac{60}{49}} f$, or $\frac{\sigma f}{\mu_0}\sim 1$ then $\mbox{BR}(A \to t\bar{t})$
tends to be the second largest branching ratio whereas $\mbox{BR}(A \to gg)$ and $\mbox{BR}(A \to WW)$ are the third
and fourth largest, respectively. Anew the branching fraction of the decay $A\to ZZ$ is smaller than $\mbox{BR}(A \to WW)$,
can be comparable with $\mbox{BR}(A \to t\bar{t})$ and is bigger than $\mbox{BR}(A \to \gamma\gamma)$
whilst the branching ratio of the decay $A\to \gamma\gamma$ is somewhat larger than $\mbox{BR}(A \to \gamma Z)$.
With increasing $\mu_0$ the branching fractions of the decays $A\to gg, WW, ZZ, \gamma\gamma$ and $\gamma Z$
reduce. In the scenarios under consideration the branching fractions $\mbox{BR}(A \to gg)$, $\mbox{BR}(A \to WW)$ and
$\mbox{BR}(A \to \gamma\gamma)$ decreases from $10^{-3}$, $10^{-4}$ and $10^{-5}$ to $10^{-4}$, $10^{-5}$
and $10^{-6}$, respectively, if $\mu_0$ changes from $1\,\mbox{TeV}$ to $4\,\mbox{TeV}$. $\mbox{BR}(A \to \eta\bar{\eta})$
is always extremely close to unity as long as $\lambda_{\eta}\gtrsim 1$. The branching ratio associated with the decay $A\to t\bar{t}$
also does not change much when $\mu_0$ varies. It remains much smaller than $\mbox{BR}(A \to \eta\bar{\eta})$ because
the corresponding decay width is suppressed by the ratio $(m_t/\Lambda_t)^2$.

At the same time the LHC production cross sections of the pseudoscalar $A$ in the scenarios with narrow and large widths
are basically the same (see Figs.~2b and 2d). In both cases $\sigma_A$ is basically determined by the masses of the exotic
quarks and their couplings to the SM singlet pNGB state $A$. Since $A$ decays predominantly into a pair $\eta\bar{\eta}$
that gives rise to the $E^{miss}_T$ in the final state, the large LHC production cross section of the pseudoscalar $A$
should also result in a non-negligible cross section for $pp\to j + E^{miss}_{T}$ that may be observed at the LHC in the
near future\footnote{In the E$_6$CHM the monojet cross-section tends to be suppressed at least by a factor $(\alpha_s/\pi)$
as compared to the LHC production cross section of the SM singlet pNGB state $A$. }.

The presence of exotic fermions with masses in the few TeV range should lead to remarkable signatures which have already
been mentioned in \cite{Nevzorov:2015sha}. The production processes of exotic coloured fermions involve gluon--induced
QCD interactions, so that these states are doubly produced. Assuming that such exotic coloured states couple most strongly to
the third generation fermions, they decay into a pair of third generation quarks and the lightest neutral exotic state
resulting in the enhancement of the cross sections for
\be
pp\to t\overline{t}b\overline{b} + E^{miss}_{T}+X \qquad \mbox{and} \qquad
pp\to b\overline{b}b\overline{b} +  E^{miss}_{T}+X\,.
\label{23}
\ee
Although the production cross sections of other exotic fermions tend to be somewhat smaller, their pair production can also
result in an enhancement of the cross sections for the processes with the final states (\ref{23}). Thus the exotic fermion states
should be observable in dedicated searches at Run 2 of the LHC.

The collider signatures associated with the scalar colour triplet $T$, that
stems from the same pNGB $SU(5)$ multiplet as the Higgs doublet,
are model--dependent. In general $T$ decays as \cite{Nevzorov:2015sha}
$$
T\to b + \overline{\zeta}_1 + X\,,
$$
where $\zeta_1$ is the lightest exotic fermion state with baryon number $B_{\zeta_1}=1/3$.
However, when the composite fermion $D^c$ carries baryon number $B_{D^c}=1/3$
the following decay channels are also allowed for this exotic scalar state $T$ \cite{Nevzorov:2015sha}
$$
T \to \overline{t}+\overline{b}+\zeta_1+\zeta_1+X\,,\qquad
T \to \overline{t} + t + b +\overline{\zeta}_1 +X\,.
$$
If $T$ is sufficiently light the scalar colour triplets can be pair--produced at the LHC.
Then the $T\overline{T}$ production can lead to the enhancement of the cross sections
associated with the channels (\ref{23}) and/or even to the enhancement of the cross sections
for the processes with six third generation quarks in the final states, i.e.
$$
pp \to  T\overline{T}\to t\overline{t}t\overline{t}b\overline{b}+E^{miss}_{T}+X\,,\qquad
pp \to  T\overline{T}\to b\overline{b}b\overline{b}b\overline{b}+E^{miss}_{T}+X\,.
$$

\section{Conclusions}

In the $E_6$ inspired composite Higgs model (E$_6$CHM) the strongly interacting sector
possesses an $SU(6)\times U(1)_B\times U(1)_L$ global symmetry. Global $U(1)_B$ and
$U(1)_L$ symmetries ensure the conservation of baryon and lepton numbers.
Near the scale $f\gtrsim 10\,\mbox{TeV}$ the $SU(6)$ global symmetry is broken
down to its $SU(5)$ subgroup, that contains the SM gauge group. Such breakdown of
the $SU(6)$ symmetry leads to eleven pNGB states.
This set of pNGB states involves, in particular, the SM--like Higgs doublet and
SM singlet boson $A$. If CP is conserved then the SM singlet scalar $A$ is CP--odd and its
mixing with the SM--like Higgs boson is forbidden. Below scale $f$ the particle spectrum
of the E$_6$CHM should also include extra matter beyond the SM  that leads to the approximate
unification of the SM gauge couplings and gives rise to a set of exotic vector--like fermions
as well as a composite, right-handed top quark. Baryon number conservation guarantees
that the lightest exotic fermion state $\eta$ is stable. Because of this the phenomenological viability
of the model under consideration requires the lightest exotic fermion to be a SM singlet,
so that it can play the role of dark matter.

In this paper we studied the possible collider signatures associated with the presence of relatively
light pseudoscalar $A$ that might be observed at the LHC in the near future. This SM singlet
pNGB state $A$ is expected to have a mass $m_A$ which is substantially lower than $f$.
In general the exotic fermions tend to gain masses which are much larger than $m_A$. Nevertheless
we assumed that the mass of the lightest exotic fermion $\mu_{\eta}$ is rather close to $m_A/2$
and its Yukawa coupling to the SM singlet pNGB state $\lambda_{\eta}\gtrsim 1$. 
As a consequence, the annihilation cross section for $\eta\, \bar{\eta} \to \mbox{SM particles}$
is sufficiently large. In this case the relic abundance of the lightest exotic fermion state
is determined by the mechanism that generates the baryon asymmetry of the Universe.
We allowed $\mu_{\eta}$ to be either smaller or bigger than $m_A/2$. The interactions of charged exotic
fermions with $A$ induce couplings of this SM singlet pNGB state to the SM gauge bosons. We considered
the interactions of $A$ with exotic fermions and specified the couplings of this pseudoscalar boson to
$gg$, $\gamma\gamma$, $W W$, $ZZ$ and $Z\gamma$.

At the LHC the pseudoscalar $A$ is mostly produced through gluon fusion. In our analysis we focused
on the region of the E$_6$CHM parameter space where the LHC production cross section of the
SM singlet pNGB state $A$ could be strongly enhanced. In this part of the parameter space all charged
exotic fermions have masses in the few TeV range and their couplings to $A$ are either larger than or
of the order of unity. We argued that in this case for $m_A\simeq 750\,\mbox{GeV}$ the LHC production
cross section $\sigma_A$ of the pseudoscalar $A$ can be as large as $0.1-1\,\mbox{pb}$.
At the same time the requirement of naturalness implies that the LHC production
cross section of the SM singlet pNGB state $A$ could vary from $0.01\,\mbox{fb}$
to a few fb. Such small values of the cross section $\sigma_A$ makes
the observation of the corresponding pseudoscalar state rather problematic. 
When $\mu_{\eta} > m_A/2$ the SM singlet pNGB state $A$ decays
predominantly into either $t\bar{t}$ or $gg$. $\mbox{BR}(A \to WW)$ tends to be considerably smaller
than $\mbox{BR}(A \to gg)$. The branching fraction of the decay $A\to ZZ$ is less than $\mbox{BR}(A \to WW)$
but bigger than $\mbox{BR}(A \to \gamma\gamma)$ while the branching ratio of the decay $A \to \gamma\gamma$
is somewhat larger than $\mbox{BR}(A \to \gamma Z)$. The qualitative pattern of branching ratios
mentioned above may vary if, for example, the masses of exotic fermions are very different.
In particular, we identified the part of the E$_6$CHM parameter space where $\mbox{BR}(A \to \gamma\gamma)$
may be the third largest branching fraction and can be as large as $10\%$ whereas the decay rates of $A\to WW, ZZ$
and $\gamma Z$ are suppressed\footnote{The obtained diphoton decay rate satisfies theoretical constraints
discussed in \cite{Low:2015qho}}. The decay channel $A\to \gamma\gamma$ may play an important role in
searches for such SM singlet pNGB state because the decay modes $A\to gg,\,t\bar{t},\,WW$ and $ZZ$ result in
final states that predominantly contain jets and therefore their experimental identification can be rather problematic.
In all these scenarios the total width of the pseudoscalar $A$ is rather narrow, i.e. $\Gamma_A/m_A \sim 10^{-5}-10^{-4}$.

When $\mu_{\eta} < m_A/2$ the tree-level decays of $A$ into $\eta\, \bar{\eta}$ are kinematically allowed
and the total width of the SM singlet pNGB state $A$ can be quite large, i.e. $\Gamma_A/m_A \gtrsim 0.01$.
In this case the pseudoscalar $A$ decays mainly into $\eta\, \bar{\eta}$ resulting in invisible final states.
If $\lambda_{\eta}\gtrsim 1$ all other branching ratios are extremely small. $\mbox{BR}(A \to gg)$
and $\mbox{BR}(A \to t\bar{t})$ can be either the second or third largest branching fractions and tend
to be less than $10^{-4}-10^{-3}$. On the other hand the LHC production cross section $\sigma_A$ of the
pseudoscalar $A$ is set by the masses of the exotic quarks as well as their couplings to $A$. Therefore
$\sigma_A$ remains basically the same as in the case of $\mu_{\eta} > m_A/2$. Then sufficiently large
$\sigma_A$ should also lead to a non-negligible cross section of the process $pp\to j + E^{miss}_{T}$.

Finally, the presence of exotic coloured fermions with masses in the few TeV range should give rise to
an enhancement of the cross sections $pp\to t\overline{t}b\overline{b} + E^{miss}_{T}+X $ and
$pp\to b\overline{b}b\overline{b} +  E^{miss}_{T}+X$, which may be also observable at Run 2 of the LHC.

%In this case one charged exotic fermion, that does not carry colour charge, and the lightest SM singlet
%exotic fermion have masses just above $375\,\mbox{GeV}$, whereas all other exotic states can
%be considerably heavier, lying even beyond the projected LHC sensitivity.

\section*{Acknowledgements}
RN is grateful to  E.~Boos,  S.~F.~King,  M.~M\"{u}hlleitner, D.~G.~Sutherland, X.~Tata, M.~Vysotsky for helpful discussions.
RN also thanks H.~Fritzsch for very valuable comments and remarks. This work was supported by the University of Adelaide
and the Australian Research Council through the ARC Center of Excellence in Particle Physics at the Terascale (CE 110001004)
and through grant LF099 2247 (AWT).

\end{document}